# Switching of a Single Photon by Two Λ-type Three-Level Quantum Dots Embedded in Cavities Coupling to One-Dimensional Waveguide

Nam-Chol Kim,[*]  Myong-Chol Ko

Department of Physics, **Kim Il Sung** University, Pyongyang, D P R Korea

**Abstract:** Switching of a single photon interacting with two Λ-type three-level quantum dots embedded in cavities coupled to one-dimensional waveguide is investigated theoretically via the real-space approach. We demonstrated that switching of a single photon can be achieved by tuning the classic driving field on or off, and by controlling the QD-cavity coupling strength, Rabi frequency and the cavity-waveguide coupling rate. The transmission properties of a single photon by such a nanosystem discussed here could find the applications in the design of next-generation quantum devices and quantum information.

[*] Electronic mail:  ryongnam10@yahoo.com

## 1. Introduction

The interaction between light and matter has always been an important topic in physics for some fundamental investigations of photon-atom interaction and for its applications in quantum information, and its most elementary level is the interaction between a single photon and a single emitter [1, 2]. Recently, controlling single photon transport is a central topic in quantum information and optical devices and theoretical idea of a single photon transistor has also emerged [3].

Many theoretical [4-7] and experimental[8-10] works have been reported for the photon scattering in different quantum systems. Most proposals for a single photon transport are based on the real-space method [4, 5], and they have mainly considered the scattering properties of a single photon interacting with emitters which have two-level structure [4-7, 11-14]. Recently, scattering properties of a single-photon by a multi-level system has been reported [3, 15-19]. The multi-level system coupled to cavity has also been widely investigated in Refs [19, 20], which showed that a strong modification of photon transmission spectra could be achieved. However, they mainly focus on the case where there exsists only a single quantum emitter with multi-level energy structure.

Here, we investigate theoretically the scattering properties of a single photon interacting with two Λ-type quantum dots, each of which is embedded in cavities, respectively, coupling to one-dimensional waveguide.

## 2. Theoretical model and dynamics equations

The schematic diagram of the system considered in this paper is exhibited in Fig. 1(a), where the two Λ-type quantum dots are embedded in cavities coupling to one-dimensional waveguide, respectively. $\omega$ is the angular frequency of incident photon field in 1D waveguide. From the dispersion relation of a single-mode waveguide, $\omega$ can be expressed as $\omega = \omega_0 + v_g k$, where $\omega_0$ is an arbitrary frequency that is away from the cutoff of the dispersion, and $k$ is the wave vector and $v_g$ the group velocity corresponding to $\omega$ [5]. Figure 1(b) shows the energy-level configuration of the Λ-type quantum dots considered in this paper. |1⟩ and |2⟩ are coupled with cavity mode and the coupling strength is $g_k$. |3⟩ and |2⟩ are coupled by a classic optical field with Rabi frequency Ω. The

Hamiltonian of the system composed of one-dimensional waveguide, two cavities, two Λ-type QDs and reservoir is given in real space by [4, 5]

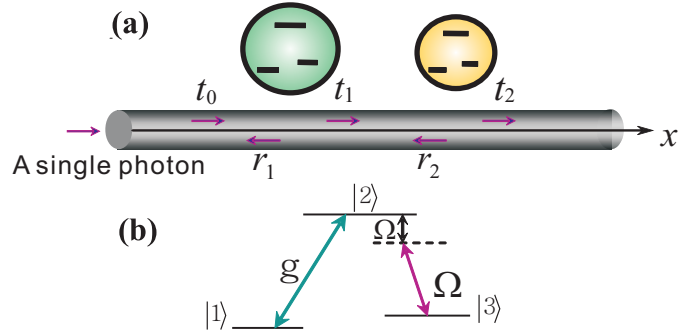

Fig.1 (Color online). (a) Schematic diagram of a system consisting of two Λ-type QDs coupled to one-dimensional waveguide, where the two Λ-type QDs are embedded in cavities, respectively. $t_k$ and $r_k$ are the transmission and reflection amplitudes at the place of $k$th QDs, $x_k$, respectively. (b) Energy configuration of the Λ-type QD.

$$H/\hbar = \int dx \left[ c_R^+(x)\left(\omega_0 - iv_g \frac{\partial}{\partial x}\right)c_R(x) + c_L^+(x)\left(\omega_0 + iv_g \frac{\partial}{\partial x}\right)c_L(x) \right]$$

$$+ \sum_{k=1}^{2}\left[\left(\omega_c^{(k)} - i\frac{1}{\tau_c^{(k)}}\right)a_k^+ a_k + \left(\omega_2^{(k)} - i\frac{1}{\tau_2^{(k)}}\right)\sigma_{22}^{(k)} + \left(\omega_2^{(k)} - \Delta^{(k)} - i\frac{1}{\tau_3^{(k)}}\right)\sigma_{33}^{(k)} + \omega_1^{(k)}\sigma_{11}^{(k)}\right] \quad (1)$$

$$+ \sum_{k=1}^{2}\left[g_k\left(a_k\sigma_{21}^{(k)} + a_k^+\sigma_{12}^{(k)}\right) + \frac{\Omega_k}{2}\left(\sigma_{23}^{(k)} + \sigma_{32}^{(k)}\right)\right]$$

$$+ \sum_{k=1}^{2} \int dx V_k \delta(x_k)\left[c_R^+(x)a_k + a_k^+ c_R(x) + c_L^+(x)a_k + a_k^+ c_L(x)\right]$$

In Eq. (1), $c_R^+(x)[c_R(x)]$ means creating (annihilating) a right-going photon at position $x$, and $c_L^+(x)[c_L(x)]$ means creating (annihilating) a left-going photon at position $x$. $\omega_c^{(k)}$ is the angular frequency of photon in the $k$ th cavity. $a_k^+(a_k)$ means creating (annihilating) a photon in the $k$ th cavity. $\omega_{21}^{(k)} = \omega_2^{(k)} - \omega_1^{(k)}$ is the transition frequency between state $|1\rangle$ and $|2\rangle$ of the $k$ th QD. $1/\tau_c^{(k)}, 1/\tau_2^{(k)}, 1/\tau_3^{(k)}$ are the dissipation rates of the $k$ th cavity and state $|2\rangle$ and $|3\rangle$ of the $k$ th QD, respectively. $\Delta^{(k)}$ is the detuning between energy of classical optical field and transition energy between states $|2\rangle$ and $|3\rangle$ of the $k$ th QD. $\sigma_{ij}^{(k)}$ is the dipole transition operator between the states $|i\rangle$ and $|j\rangle$ of the $k$ th QD. The coupling rate between $k$ th cavity and waveguide is denoted by $V_k$.

Assuming that a single photon is incoming from the left with energy $E_k = \hbar\omega$, then the eigenstate of the system, defined by $H|\psi\rangle = E_w|\psi\rangle$, can be constructed in the form

$$|\psi\rangle = \int dx[\phi_R(x)c_R^+(x) + \phi_L(x)c_L^+(x)]|0,0,1\rangle + \sum_{k=1}^{2}[c_c^{(k)}a_k^+|0,0,1\rangle + c_2^{(k)}|0,0,2\rangle + c_3^{(k)}|0,0,3\rangle], \quad (2)$$

where $E_w = \hbar\left(\omega + \sum_{k=1}^{2}\omega_1^{(k)}\right)$ is the energy eigenvalue of the whole system. $\phi_{R/L}(x)$ is the single photon wave function in the $R/L$ mode at $x$. $c_c^{(k)}$ is the excitation amplitude of the $k$ th cavity and $c_i^{(k)}$ is the excitation amplitude of the state $|i\rangle$ $(i = 2, 3)$ of the $k$ th QD. Here, $|0, 0, i\rangle$ indicates the waveguide and cavity in vacuum state and QD in state $|i\rangle$ $(i = 1, 2, 3)$. For a single photon incident from the left, the mode functions $\phi_R^+(x)$ and $\phi_L^+(x)$ take the forms $\phi_R^+(x < 0) = e^{ikx}$, $\phi_R^+(0 < x < l) = t_1 e^{ik(x-l)}$, $\phi_R^+(x > l) = t_2 e^{ik(x-2l)}$, $\phi_L^+(x < 0) = r_1 e^{-ikx}$, $\phi_L^+(0 < x < l) = r_2 e^{-ik(x-l)}$, and $\phi_L^+(x > l) = 0$, respectively, where $l$ is the spacing between the two QDs. Here $t_k$ and $r_k$ are the transmission and reflection amplitudes at the place $x_k$, respectively.

Substituting Eqs (1) and (2) into the Schrödinger equation, $H|\psi\rangle = E_w|\psi\rangle$, we obtain the following equations, $-iv_g(t_1 e^{-ikl} - t_0) + V_1 c_c^{(1)} = 0$, $iv_g(r_2 e^{ikl} - r_1) + V_1 c_c^{(1)} = 0$, $-C_1 c_c^{(1)} + g_1 c_2^{(1)} + V_1\left[\frac{1}{2}(t_0 + t_1 e^{-ikl}) + \frac{1}{2}(r_1 + r_2 e^{ikl})\right] = 0$, $-A_1 c_2^{(1)} + c_c^{(1)} g_1 + \frac{\Omega_1}{2} c_3^{(1)} = 0$,

$-c_3^{(1)}(B_1 + \Delta^{(1)}) + \frac{\Omega_1}{2} c_2^{(1)} = 0$, $-iv_g(t_2 e^{-ikl} - t_1) + V_1 c_c^{(2)} = 0$, $iv_g(r_3 e^{ikl} - r_2) + V_2 c_c^{(2)} = 0$,

$-C_2 c_c^{(2)} + g_2 c_2^{(2)} + V_2\left[\frac{1}{2}(t_1 + t_2 e^{-ikl}) + \frac{1}{2}(r_2 + r_3 e^{ikl})\right] = 0$, $-A_2 c_2^{(2)} + c_c^{(2)} g_2 + \frac{\Omega_2}{2} c_3^{(2)} = 0$,

$-c_3^{(2)}(B_2 + \Delta^{(2)}) + \frac{\Omega_2}{2} c_2^{(2)} = 0$. Here $A_k = (\omega - \omega_{21}^{(k)}) + \frac{i}{\tau_2^{(k)}}$, $B_k = (\omega - \omega_{21}^{(k)}) + \frac{i}{\tau_3^{(k)}}$,

$C_k = (\omega - \omega_c^{(k)}) + \frac{i}{\tau_c^{(k)}}$, where $k = 1, 2$. By solving the above set of equations, we can obtain the transmission coefficient and reflection coefficient exhibiting the scattering properties of an incident single photon by two Λ-type quantum dot systems in cavities coupled to one-dimensional waveguide, respectively.

**3. Theoretical analysis and numerical results**

## 3. 1. Transmission of a single photon versus incident frequency of a single photon

### 3. 1. 1. QD-classical optical field in tune and the classic driving field tuning off case ($\Delta_1 = \Delta_2 = 0$; $\Omega_1 = \Omega_2 = 0$).

For the QD and classical optical field in tune and the classic driving field tuning off case, each $\Lambda$-type three-level structure of quantum dots degenerates to single two-level system. The scattering property of the single photon in the long time limit is characterized by the transmission coefficient $T_2 \equiv |t_2|^2$ and reflection coefficient $R_1 \equiv |r_1|^2$, where the transmission and reflection amplitudes are given, respectively, as

$$\begin{cases} t_2 = \dfrac{-(AC-g^2)^2}{2A^2J^2 - 2iAJ(AC-g^2) - (AC-g^2)^2}, \\ r_1 = -\dfrac{2A^2J^2}{2A^2J^2 - 2iAJ(AC-g^2) - (AC-g^2)^2}. \end{cases} \quad (3)$$

We first consider the case the QDs and cavities are tuned, respectively, i. e., $\delta_1 = \delta_2 = 0$, where $\delta_k = \omega_{21}^{(k)} - \omega_c^{(k)}$, $k = 1, 2$. In all our calculations, we suppose that $1/\tau_c^{(k)} = 0$, $1/\tau_2^{(k)} = 0$, $1/\tau_3^{(k)} = 0$, $V_1 = V_2$, $\omega_{21}^{(1)} = \omega_{21}^{(2)}$, $\omega_c^{(1)} = \omega_c^{(2)}$, therefore $J = J_k = V_k^2/v_g$, $A = A_k$, $B = B_k$, $C = C_k$ ($k = 1, 2$). The transmission coefficient $T_2$ equals to maximum 1 at $\omega = \omega_{21}$ and equals to minimum 0 at $\omega = \omega_{21} \pm g$ as shown in Fig. 2(a), which shows that the incident photon can be transmitted completely at resonance, $\omega = \omega_{21}$, and reflected completely at $\omega = \omega_{21} \pm g$, which are quite different with the scattering spectra in Ref. [12]. From the results obtained here, we can see the role of the existence of cavities resulting in changes of the scattering of an incident single photon. In the calculations, we set the parameters $g_1 = g_2 = 1.3 \times 10^{-5} \omega_{21}$, $J_1 = J_2 = 1.4 \times 10^{-5} \omega_{21}$, which was reported in a system with a single quantum dot coupled to a cavity in Ref. [21]. When $g$ is small, $g = g_1 = g_2 = 0.3 \times 10^{-5} \omega_{21}$, the width of the transmission peak becomes narrow. When the QDs and the cavities are detuned, respectively, the transmission spectra are shown in Fig. 2(b). The transmission coefficient $T_2$ equals to maximum 1 at $\omega = \omega_{21}$ and equals to minimum 0 at $\omega = \omega_{21} - \left(\sqrt{4g^2 + \delta^2}/2 + \delta/2\right)$ and $\omega = \omega_{21} + \left(\sqrt{4g^2 + \delta^2}/2 - \delta/2\right)$, respectively. When $g$ is small, the width of transmission peak also becomes narrow as shown with the solid line in Fig. 2(b) and the complete reflection can be achieved near the resonance $\omega \approx \omega_{21}$.

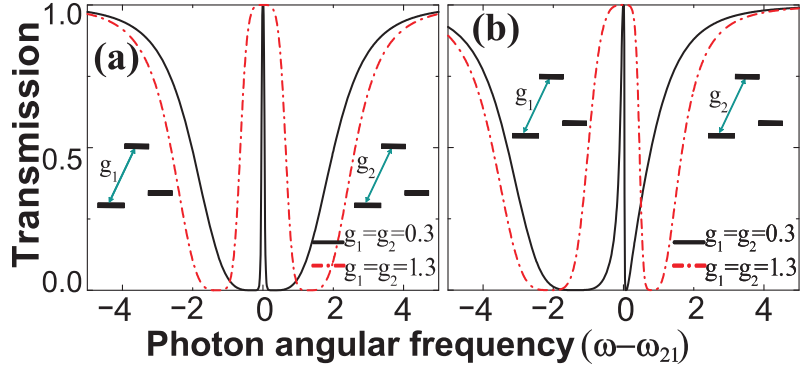

Fig. 2 (Color online). Single-photon transmission spectrum with no detuning ($\Delta_1=\Delta_2=0$) for two-level system in the QD and cavity tuned ($\delta_1=\delta_2=0$) (a) and detuned ($\delta_1=\delta_2=1.3\times10^{-5}\omega_{21}$) (b) cases. (a) $g_1=g_2=1.3\times10^{-5}\omega_{21}$ (dash-dotted line); $g_1=g_2=0.3\times10^{-5}\omega_{21}$(solid line). (b) $g_1=g_2=1.3\times10^{-5}\omega_{21}$(dash-dotted line); $g_1=g_2=0.3\times10^{-5}\omega_{21}$(solid line). Insets show the energy-level couplings of QD with classic field tuning off (a, b). Here, $\Omega_1=\Omega_2=0$, $J_1=J_2=1.4\times10^{-5}\omega_{21}$, $l=\lambda/4+n\lambda/2$ ($n=0, 1, 2, \ldots$) and the frequency is in unit of $10^{-5}\omega_{21}$.

### 3. 1. 2. QD-classical optical field in tune and the classic driving field tuning on case ($\Delta_1=\Delta_2=0$; $\Omega_1=\Omega_2\neq 0$).

For the QD and classical optical field in tune and the classic driving field tuning on case, the transmission and reflection amplitudes are given, respectively, as

$$\begin{cases} t_2 = -\dfrac{(4ABC-4Bg^2-C\Omega^2)^2}{2(4AB-\Omega^2)^2 J^2 - 2iJ(4AB-\Omega^2)(4ABC-4Bg^2-C\Omega^2)-(4ABC-4Bg^2-C\Omega^2)^2}, \\ r_1 = -\dfrac{2(4AB-\Omega^2)^2 J^2}{2(4AB-\Omega^2)^2 J^2 - 2iJ(4AB-\Omega^2)(4ABC-4Bg^2-C\Omega^2)-(4ABC-4Bg^2-C\Omega^2)^2}. \end{cases} \quad (4)$$

In the QDs and cavities tuned cases, respectively, the transmission spectrum is shown in Fig. 3(a), which shows that the transmission coefficient $T_2$ equals to maximum 1 at $\omega=\omega_{21}\pm\Omega/2$ and equals to minimum 0 at $\omega=\omega_{21}$, $\omega_{21}\pm\sqrt{g^2+\Omega^2/4}$. One can find that the width of transmission peak becomes wide as $g$ becomes large and the position of the peak depends only on the Rabi frequency $\Omega$. What is more interesting is that there are two complete transmission peaks appear near the resonance, which is quite different with the result for two two-level QDs system reported in Ref. [12]. For the two QDs and cavities detuned cases, respectively, the transmission spectrum is shown in Fig.

3(b). From Eq. (4), one can find the incident photon is completely transmitted at $\omega = \omega_{21} \pm \Omega/2$, but the shape of the curve is asymmetrical with respect to $\omega = \omega_{21}$ due to detuning δ as shown in Fig. 3(b). The position of complete reflection peak can be determined with the condition $4ABC - 4Bg^2 - C\Omega^2 = 0$.

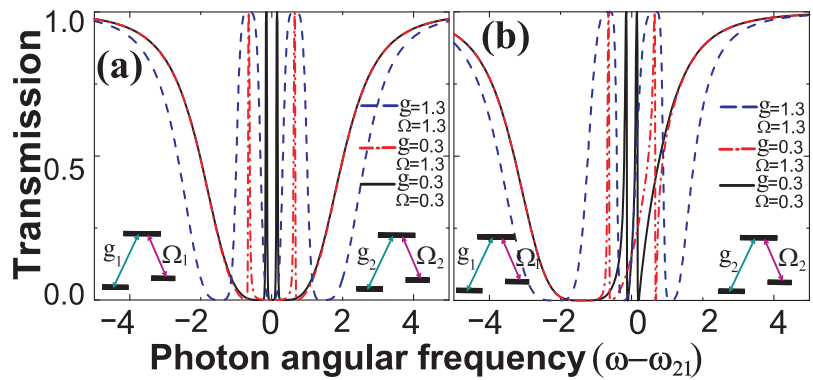

Fig. 3 (Color online). Single-photon transmission spectrum with no detuning ($\Delta_1=\Delta_2=0$) for three-level system in the QD and cavity tuned ($\delta_1= \delta_2 =0$) (a) and detuned ($\delta_1= \delta_2 =1.3\times10^{-5}\omega_{21}$) (b) cases. Insets show the energy-level couplings of atom with classic field tuning on (a, b). The parameters are taken $g_1=g_2=g=1.3\times10^{-5}\omega_{21}$, $\Omega_1=\Omega_2=\Omega=1.3\times10^{-5} \omega_{21}$,(dashed line); $g_1=g_2=g=0.3\times10^{-5}\omega_{21}$, $\Omega_1=\Omega_2=\Omega=1.3\times10^{-5}\omega_{21}$ (dash-dotted line); $g_1=g_2=g=0.3\times10^{-5}\omega_{21}$, $\Omega_1= \Omega_2=\Omega=0.3\times10^{-5}\omega_{21}$ (solid line). Here $J_1=J_2=1.4\times10^{-5}\omega_{21}$, $l = \lambda/4+n \lambda/2$ ($n$ =0, 1, 2, …) and the frequency is in unit of $10^{-5}$ $\omega_{21}$.

### 3. 1. 3. QD-classical optical field detuned and the classic driving field tuning on case ($\Delta_1= \Delta_2= \Delta\neq0$; $\Omega_1= \Omega_2= \Omega\neq0$).

We can also consider the QD and classical optical field detuned and the classic driving field tuning on cases. When $\Delta_1=\Delta_2=\Delta\neq0$ and $\Omega_1=\Omega_2=\Omega\neq0$, $T_2$ becomes asymmetric with respect to $\omega = \omega_{21}$ as shown in Fig. 4 (a) and 4(b). $T_2$ equals to maximum 1 at $\omega_{21} - \left(\sqrt{\Delta^2 + \Omega^2} + \Delta\right)/2$ and $\omega_{21} + \left(\sqrt{\Delta^2 + \Omega^2} - \Delta\right)/2$. For the detuning given as $\Delta_1= \Delta_2= \Delta=1.3\times10^{-5}\omega_{21}$, the smaller $g$ is, the narrower the width of the transmission peak (dashed line) as shown in Fig. 4 (a). Especially, when $g$ is large, there coexist two quite different transmission peaks, one of them is wide and the another is very narrow, which is quite similar to the results reported in Ref. [22], where $n$-1 two-level QDs have the same transition frequencies and a two-level QD has a different one. Fig. 4(b) shows the transmission spectrum with different detunigs Δ. The spacing between the two transmission peaks is $\sqrt{\Delta^2+\Omega^2}$, which shows that one can control the transmission properties by adjusting the Rabi frequencies and the detunings.

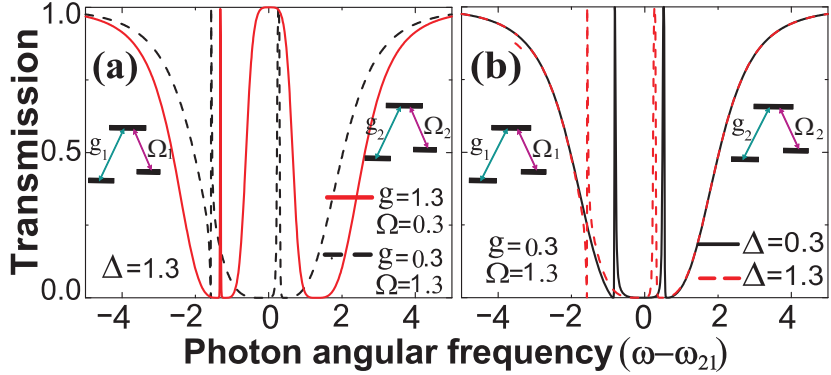

Fig. 4 (Color online). Single-photon transmission spectra with detuning ($\Delta_1=\Delta_2\neq 0$) in the QD and cavity tuned ($\delta_1=\delta_2=0$) cases (a, b). (a) $\Delta_1=\Delta_2=1.3\times10^{-5}\omega_{21}$, where g=1.3×10$^{-5}$ $\omega_{21}$, $\Omega$=0.3×10$^{-5}\omega_{21}$(dashed line), g=0.3×10$^{-5}\omega_{21}$, $\Omega$=1.3×10$^{-5}\omega_{21}$ (solid line). (b) g=0.3×10$^{-5}$ $\omega_{21}$, $\Omega$=1.3×10$^{-5}\omega_{21}$, where $\Delta_1=\Delta_2$=1.3×10$^{-5}\omega_{21}$ (dashed line), $\Delta_1=\Delta_2$=0.3×10$^{-5}$ $\omega_{21}$ (solid line). Insets show the energy-level couplings of atom with classic field tuning on (a, b). Here $J_1=J_2$=1.4×10$^{-5}\omega_{21}$, $l=\lambda/4+n\lambda/2$ ($n$ =0, 1, 2, …) and the frequency is in unit of 10$^{-5}$ $\omega_{21}$.

### 3. 1. 4. QD-classical optical field in tune case with different coupling strengths between cavity and waveguide ($\Delta_1=\Delta_2=\Delta=0$; $J_1=J_2=J$=1.4·10$^{-5}$ $\omega_{21}$, 5·10$^{-5}$ $\omega_{21}$, 14·10$^{-5}$ $\omega_{21}$)

We can also investigate the influence of the coupling strength between $k$ th cavity and waveguide $J=J_k=V_k^2/v_g$ in tune case. For the energy-level couplings of QDs with classic field tuning off ($\Omega_1=\Omega_2=0$), the transmission coefficient $T_2$ is obtained with

$$\begin{cases} t_2 = \dfrac{-(AC-g^2)^2}{2A^2J^2-2iAJ(AC-g^2)-(AC-g^2)^2}, \\ r_1 = -\dfrac{2A^2J^2}{2A^2J^2-2iAJ(AC-g^2)-(AC-g^2)^2}. \end{cases} \quad (5)$$

$T_2$ equals to maximum 1 at $\omega=\omega_{21}$ and equals to minimum 0 at $\omega=\omega_{21}\pm g$ as shown in Fig. 5(a), which shows there exists only one complete transmission peak at resonance $\omega=\omega_{21}$ and the transmission spectrum becomes narrow when the coupling $J$ becomes large. However, for the energy-level couplings of QDs with classic field tuning on ($\Omega_1=\Omega_2\neq 0$) case, $T_2$ equals to maximum 1 at $\omega=\omega_{21}\pm\Omega/2$ and equals to minimum 0 at $\omega=\omega_{21}$, $\omega_{21}\pm\sqrt{g^2+\Omega^2/4}$ as shown in Fig. 5(b). In the classic driving field tuning on case, $T_2$ equals to minimum 0 at $\omega = \omega_{21}$, which is quite different with tuning off case. This means that the incident single photon can be reflected or transmitted completely at resonance by tuning the classic driving field on or off. One can also observe that the number of transmission peaks changes from one to two by tuning the classic driving field on.

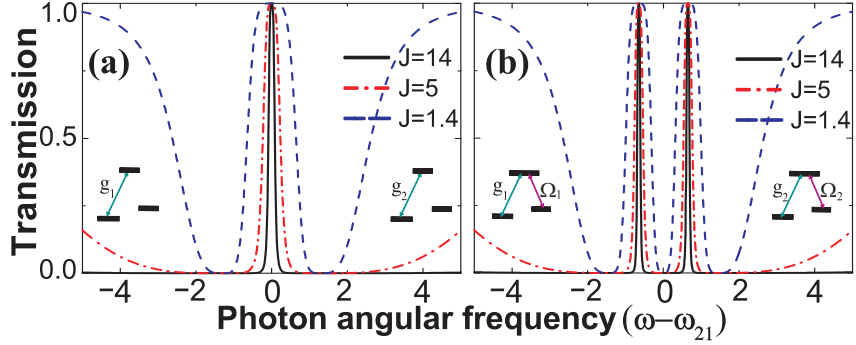

Fig. 5 (Color online). Single-photon transmission spectra with no detuning ($\Delta_1=\Delta_2=0$) in the QD and cavity tuned ($\delta_1= \delta_2 =0$) cases (a, b). (a) $J_1=J_2= J=14\times10^{-5}$ $\omega_{21}$(solid line) ; $J_1=J_2= J=5\times10^{-5}$ $\omega_{21}$(dash-dotted line); $J_1=J_2=J=1.4\times10^{-5}$ $\omega_{21}$(dashed line) (b) $J_1=J_2=J=14\times10^{-5}$ $\omega_{21}$(solid line) ; $J_1=J_2= J=5\times10^{-5}$ $\omega_{21}$(dash-dotted line); $J_1= J_2= J= 1.4\times10^{-5}\omega_{21}$ (dashed line). Insets show the energy-level couplings of atom with classic field tuning off ($\Omega_1= \Omega_2=0$) (a) and on ($\Omega= \Omega_1= \Omega_2=1.3\times10^{-5}\omega_{21}$) (b). Here $g =1.3\times10^{-5}$ $\omega_{21}$, $l = \lambda / 4+n \lambda / 2$( $n =0, 1, 2, \ldots$) and the frequency is in unit of $10^{-5}$ $\omega_{21}$.

### *3. 1. 5. QD-classical optical field in tune case with different coupling strengths between cavities and QDs and different Rabi frequencies ($\Delta_1= \Delta_2= \Delta=0$, $\Omega_1\neq0$, $\Omega_2\neq0$).*

We can also obtain the single-photon transmission spectra with no detuning ($\Delta_1=\Delta_2=0$) in the QD and cavity tuned ($\delta_1= \delta_2 =0$) cases as shown in Figs. 6 (a) and 6 (b). As shown in Fig. 6(a), the transmission equals to maximum 1 at $\omega_{21}$ and $\omega_{21} \pm \sqrt{g_1^2 + g_2^2}$, and equals to minimum 0 at $\omega_{21} \pm g_1$ and $\omega_{21} \pm g_2$ when $\Omega_1=\Omega_2=0$. When $\Omega_1=1.3\times10^{-5}$ $\omega_{21}$ and $\Omega_2=2.0\times10^{-5}$ $\omega_{21}$, the transmission equals to minimum 0 at $\omega_{21}$, $\omega_{21} \pm \sqrt{g^2 + \Omega_1^2/4}$ and $\omega_{21} \pm \sqrt{g^2 + \Omega_2^2/4}$ as shown in Fig. 6(b). As we can see from the graphic illustration of Figs. 6(a) and 6(b), the incident single photon can be reflected or transmitted completely by tuning the classic driving field on or off.

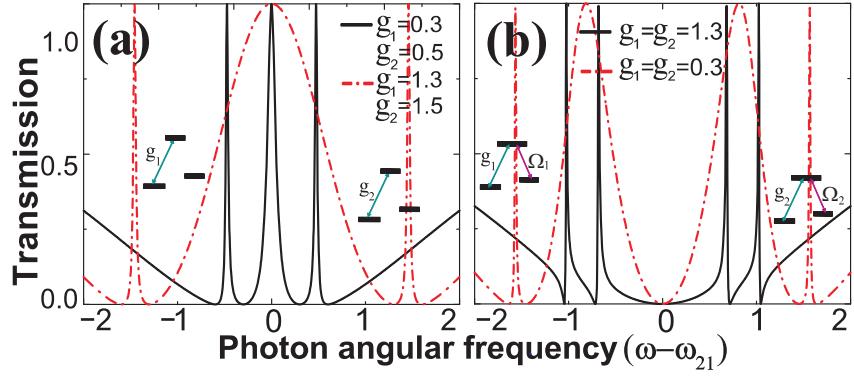

Fig. 6 (Color online). Single-photon transmission spectra with no detuning ($\Delta_1=\Delta_2=0$) in the QD and cavity tuned ($\delta_1=\delta_2=0$) cases (a, b). (a) $g_1=1.3\times10^{-5}\ \omega_{21}$; $g_2=1.5\times10^{-5}\ \omega_{21}$ (dash-dotted line); $g_1=0.3\times10^{-5}\ \omega_{21}$, $g_2=0.5\times10^{-5}\ \omega_{21}$ (solid line). (b) $g_1=g_2=1.3\times10^{-5}\ \omega_{21}$ (solid line), $g_1=g_2=0.3\times10^{-5}\ \omega_{21}$ (dash-dotted line). Insets show the energy-level couplings of QD with classic field tuning off ($\Omega_1=\Omega_2=0$) (a) and on ($\Omega_1=1.3\times10^{-5}\ \omega_{21}$, $\Omega_2=2.0\times10^{-5}\ \omega_{21}$) (b). Here $J_1=J_2=1.4\times10^{-5}\ \omega_{21}$, $l=(n+1)\lambda/2$ ($n=0, 1, 2, \ldots$) and the frequency is in unit of $10^{-5}\ \omega_{21}$.

### *3. 2. Transmission of a single photon with different interparticle distances*

Finally, we investigate the transmission properties of a single photon with different spacings between the QDs. Figure 7 shows the influence of the interparticle distance on the transmission of the incident single photon. As we can see easily from Figs. 7(a)-(d), when $\Omega=5\times10^{-5}\omega_{21}$(dashed line) and $\Omega=1.4\times10^{-5}\omega_{21}$(solid line), the transmission coefficient $T_2$ equals to minimum 0 at $\omega=\omega_{21}$ and $\omega_{21}\pm\sqrt{g^2+\Omega^2/4}$, and equals to maximum 1 at $\omega=\omega_{21}\pm\Omega/2$, regardless of the interparticle distances. Furthermore, we can also find a graphic illustration of the existence of another transmission peak for the interparticle distances $l=\lambda/8+n\lambda/2$ and $l=3\lambda/8+n\lambda/2$, as shown in Figs. 7 (a) and 7 (c), respectively. When $\Omega=14\times10^{-5}\omega_{21}$(dash-dotted line), $T_2$ equals to minimum 0 at $\omega=\omega_{21}$, regardless of the interparticle distances, and for the spacings $l=\lambda/4+n\lambda/2$ and $l=\lambda/2+n\lambda/2$, there is no a complete transmission peak, as shown in Figs. 7 (b) and 7 (d), respectively. However, for the other spacings as $l=\lambda/8+n\lambda/2$ and $l=3\lambda/8+n\lambda/2$, there appears a complete transmission peak, as shown in Figs. 7 (a) and 7 (c), respectively. When $\Delta_1=\Delta_2=0$ and $\delta_1=\delta_2=0$, one can control the transmission or reflection of a single photon by changing the Rabi frequency. We also found that the transmission spectrum can be changed greatly with the interparticle distances and exhibits an oscillatory pattern with half of the wavelength of the incident single-photon field as a period, which can be exploited to probe the separations of QDs.

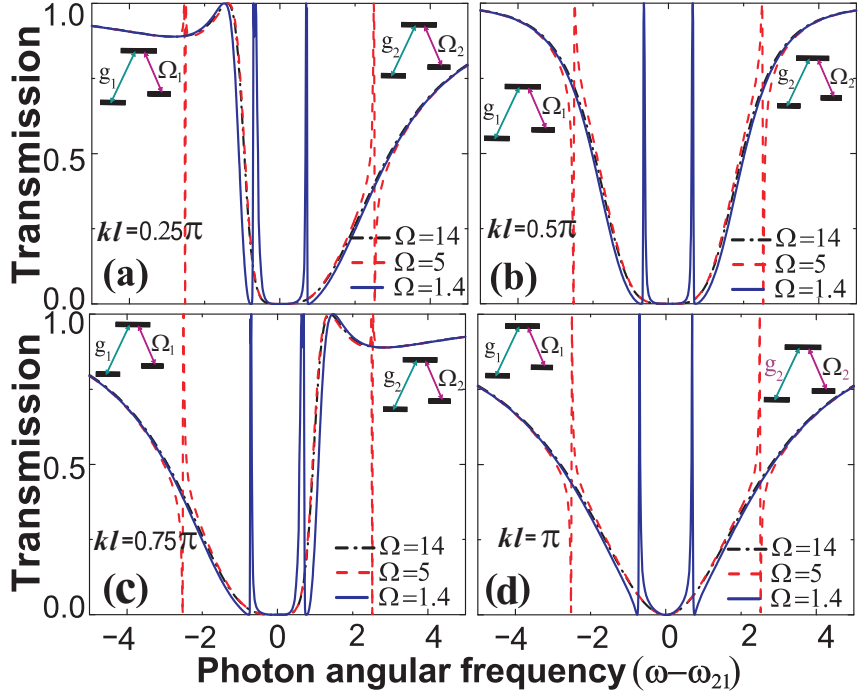

Fig. 7. Single-photon transmission spectra with no detuning ($\Delta_1=\Delta_2=0$) in the QD and cavity tuned ($\delta_1=\delta_2=0$) cases, with different spacings as (a) $l = \lambda / 8 + n \lambda / 2$, (b) $l = \lambda / 4 + n \lambda / 2$, (c) $l = 3\lambda / 8 + n \lambda / 2$, and (d) $l = \lambda / 2 + n \lambda / 2$, where $n = 0, 1, 2, \ldots$. In all cases(a, b, c, d), we set $\Omega_1 = \Omega_2 = \Omega = 14 \times 10^{-5} \omega_{21}$(dash-dotted line); $\Omega_1 = \Omega_2 = \Omega = 5 \times 10^{-5} \omega_{21}$(dashed line); $\Omega_1 = \Omega_2 = \Omega = 1.4 \times 10^{-5} \omega_{21}$(solid line). Insets show the energy-level couplings of atom with classic field tuning on. Here $g_1 = g_2 = g = 0.2 \times 10^{-5} \omega_{21}$, $J_1 = J_2 = 1.4 \times 10^{-5} \omega_{21}$ and the frequency is in unit of $10^{-5} \omega_{21}$.

## 4. Conclusions

In conclusion, we investigated theoretically the transmission spectrum of a single photon interacting with two Λ-type three-level QDs coupled to one-dimensional waveguide, where each QD is embedded in a cavity, respectively. Switching of a single photon can be achieved by tuning the classic driving field on or off and by controlling the QD-cavity coupling strength, Rabi frequency and the cavity-waveguide coupling rate. Our calculations show that the width of the transmission peak becomes narrow, when the coupling between cavity and waveguide becomes large and the coupling strength between the QD and the cavity mode becomes small, respectively. We also note that the transmission spectrum of a single photon is sensitive to the spacing between the two QDs, which can be exploited to probe the separations of quantum emitters. The results discussed in this paper could find the applications in plasmonic nanodevices and quantum information processing.

**Acknowledgments.** This work was supported by Key Project for Frontier Research on Quantum Information and Quantum Optics of Ministry of Education of D P R of Korea.

**References**
[1] K. Srinivasan and O. Painter, Nature **450** (2007) 862.
[2] G. Zumofen, N. M. Mojarad, V. Sandoghdar, M. Agio, Phys. Rev. Lett. **101** (2008) 180404.
[3] D. E. Chang, A. S. Sørensen, E. A. Demler, and M. D. Lukin, Nat. Phys. **3** (2007) 807.
[4] J. -T. Shen and S. Fan, Opt. Lett. **30** (2005) 2001.
[5] J. -T. Shen and S. Fan, Phys. Rev. A **79** (2009) 023837.
[6] D. Witthaut and A. S. Sørensen, New J. Phys. **12** (2010) 043052.
[7] M. Bradford, J. -T. Shen, Phys. Rev. A **85** (2012) 043814.
[8] B. Dayan, A. S. Parkins, T. Aoki, E. P. Ostby, K. J. Vahala, H. J. Kimble, Science **319** (2008) 1062.
[9] T. Aoki, B. Dayan, E. Wilcut, W. P. Bowen, A. S. Parkins, T. J. Kippenberg, K. J. Vahala, H. J. Kimble, Nature **443** (2006) 671.
[10] K. M. Birnbaum, A. Boca, R. Miller, A. D. Boozer, T. E. Northup, H. J. Kimble, Nature **436** (2005) 87.
[11] Yuecheng Shen and J. –T. Shen, Phys. Rev. A **85** (2012) 013801.
[12] Nam-Chol Kim, Jian-Bo Li, Zhong-Jian Yang, Zhong-Hua Hao, and Qu-Quan Wang, Appl. Phys. Lett. **97** (2010) 061110.
[13] Mu-Tian Cheng, Xiao-San Ma, Ya-Qin Luo, Pei-Zhen Wang, and Guang-Xing Zhao, Appl. Phys. Lett. **99** (2011) 223509.
[14] T. S. Tsoi, C. K. Law, Phys. Rev. A **80** (2009) 033823.

[15] Mu-Tian Cheng, Ya-Qin Luo, Pei-Zhen Wang, and Guang-Xing Zhao, Appl. Phys. Lett. **97** (2010) 191903 ; Mu-Tian Cheng, Ya-Qin Luo, Yan-Yan Song, Guang-Xing Zhao, Optics Communications, **283** (2010) 3721-3726.

[16] Z. H. Wang, Y. Li, D. L. Zhou, C. P. Sun, and P. Zhang, Phys. Rev. A **86** (2012) 023824.

[17] Wenlan Chen, Kristin M. Beck, Robert Bücker, Michael Gullans, Mikhail D. Lukin, Haruka Tanji-Suzuki, Vladan, Science, **341**, (2013) 768-770.

[18] Lukas Neumeier, Martin Leib, and Michael J. Hartmann, Phys. Rev. Lett. **111** (2013) 063601.

[19] Lan Zhou , Li-Ping Yang , Yong and C. P. Sun, Phys. Rev. Lett. **111** (2013) 103604.

[20] M. T. Cheng, X. S. Ma, M. T. Ding, Y. Q. Luo, and G. X. Zhao, Phys. Rev. A **85** (2012) 053840.

[21] T. Yoshie, A. Scherer, J. Hendrickson, G. Khitrova, H. Gibbs, G. Rupper, C. Ell, O. Shchekin, D. Deppe, Nature 432 (2004) 200.

[22] Nam-Chol Kim, Myong-Chol Ko, Song-Jin Im, Qu-Quan Wang, arXiv.1307.5643 [physics. optics]  (2013).